\documentstyle[12pt]{article}
 
\input psfig

\newcommand{\nc}{\newcommand}

\nc{\eqr}[1]{(\ref{#1})}
\nc{\sref}[1]{\S\ref{#1}}
\nc{\tref}[1]{Table~\ref{#1}}
\nc{\fref}[1]{Figure~\ref{#1}}
\nc{\cref}[1]{Chapter~\ref{#1}}
\nc{\beq}{\begin{equation}}
\nc{\eeq}{\end{equation}}
\nc{\barray}{\begin{eqnarray}}
\nc{\earray}{\end{eqnarray}}
\nc{\barrayn}{\begin{eqnarray*}}
\nc{\earrayn}{\end{eqnarray*}}
\nc{\bcenter}{\begin{center}}
\nc{\ecenter}{\end{center}}
\nc{\lra}{\longrightarrow}
\nc{\Lra}{\Longrightarrow}
\nc{\ra}{\rightarrow}
\nc{\ADE}{$A$-$D$-$E$\ }
\nc{\setall}{\setcounter{equation}{0}
        \setcounter{definition}{0}
        \setcounter{lemma}{0}
        \setcounter{convention}{0}
        \setcounter{conjecture}{0}
        \setcounter{theorem}{0}
        \setcounter{proposition}{0}
        \setcounter{property}{0}
        \setcounter{fact}{0}
        \setcounter{corollary}{0}}
\nc{\setequation}{\setcounter{equation}{0}}
\nc{\hs}[1]{\hspace{#1 mm}}

\def\sCC{{\kern 0.27em\vrule height1.45ex width0.03em depth0em
	  \kern-0.30em\rm C}}
\def\C{{\mathchoice
  {\sCC}
  {\sCC}
  {\kern 0.225em \vrule height1.05ex width0.025em depth0em \kern-0.25em \rm C}
  {\kern 0.180em \vrule height0.78ex width0.02em depth0em \kern-0.2em \rm C}
	}}
\def\sHH{{\rm I\kern-.16em{}H}}
\def\H{{\mathchoice
  {\sHH}
  {\sHH}
  {\rm I\kern-.13em{}H}
  {\rm I\kern-.13em{}H} }}
\def\sNN{{\rm I\kern-.16em{}N}}
\def\N{{\mathchoice
  {\sNN}
  {\sNN}
  {\rm I\kern-.12em{}N}
  {\rm I\kern-.10em{}N} }}
\def\sPP{{\rm I\kern-.16em{}P}}
\def\P{{\mathchoice
  {\sPP}
  {\sPP}
  {\rm I\kern-.12em{}P}
  {\rm I\kern-.10em{}P} }}
\def\sQQ{{\kern 0.27em \vrule height1.45ex width0.03em depth0em
	  \kern-0.30em \rm Q}}
\def\Q{{\mathchoice
	{\sQQ}
	{\sQQ}
  {\kern 0.225em \vrule height1.05ex width0.025em depth0em \kern-0.25em \rm Q}
  {\kern 0.180em \vrule height0.78ex width0.020em depth0em \kern-0.20em \rm Q}
	}}
\def\sRR{{\rm I\kern-0.16em{}R}}
\def\R{{\mathchoice
  {\sRR}
  {\sRR}
  {\rm I\kern-0.12em{}R}
  {\rm I\kern-0.10em{}R} }}
\def\sZZ{{\rm Z\kern-0.32em{}Z}}
\def\Z{{\mathchoice
  {\sZZ}
  {\sZZ} 
  {\rm Z\kern-0.3em{}Z}     %.3
  {\rm Z\kern-0.25em{}Z} }}  %.25
\def\ZZZ{{\rm Z\kern-0.24em{}Z}}
\def\sKK{{\rm I\kern-0.16em{}K}}
\def\K{{\mathchoice
  {\sKK}
  {\sKK}
  {\rm I\kern-0.12em{}K}
  {\rm I\kern-0.10em{}K} }}

\nc{\cznzn}{\C^3/\Z_{n+3}\times\Z_{n+3}}
\nc{\qed}{\mbox{\raisebox{-.7mm}{\large $\Box$}}}
\nc{\znzn}{\Z_{n+3}\times\Z_{n+3}}
\newtheorem{theorem}{\bf THEOREM}

\newtheorem{corollary}{\bf COROLLARY}

\renewcommand{\thefootnote}{\fnsymbol{footnote}}
\begin{document}

\begin{titlepage}
{\flushright{\small MIT-CTP-2888\\hep-th/9908008\\}}
\begin{center}

{\LARGE Three-Dimensional Gorenstein Singularities and}\\
\vspace{3mm}
{\LARGE $\widehat{SU(3)}$ Modular Invariants}
\end{center}

\vspace{1cm}
\begin{center}

{\large Jun S. Song\footnote{E-mail:
jssong@ctp.mit.edu.\\ Research supported in part
by the NSF Graduate Fellowship and the U.S. Department of Energy under
cooperative research 
agreement $\#$DE-FC02-94ER40818.
}}\\
\vspace{.8cm}
{\it Center for Theoretical Physics}\\
{\it Massachusetts Institute of Technology}\\
{\it Cambridge, Massachusetts 02139}
\end{center}

\vspace{1cm}
\begin{abstract}
Using  $N=2$  Landau-Ginzburg theories,
we examine the recent conjectures relating
the $\widehat{SU(3)}$ WZW modular invariants, finite subgroups of
$SU(3)$ and Gorenstein singularities.  
All isolated three-dimensional Gorenstein singularities do not 
appear to be related to any known Landau-Ginzburg theories, but 
we present some curious observations which suggest that the
$SU(3)_n/SU(2)\times U(1)$ Kazama-Suzuki  model may be related to a
deformed geometry of 
$\C^3/\Z_{n+3}\times\Z_{n+3}$.  The toric resolution diagrams of those
particular singularities are also seen to be  classifying
the diagonal modular invariants of the $\widehat{SU(3)}_n$ as
well as the $\widehat{SU(2)}_{n+1}$ WZW models.

\end{abstract}
\end{titlepage}

\renewcommand{\thefootnote}{\arabic{footnote}}

%%%%%%%%%%%%%%%%%%%%%%%%%%%
%	Introduction
%%%%%%%%%%%%%%%%%%%%%%%%%%%

\section{Introduction}
Study of integrable  lattice models has previously led physicists 
to speculate a possible
connection between the finite subgroups of $SU(3)$ and the modular
invariants of the $\widehat{SU(3)}$ WZW models.  In particular, it has
been
observed in \cite{DiFrancesco} that the representation theory graphs of
the $\Z_n\times\Z_n$ finite subgroups are closely related to the
diagonal 
${\cal A}$-modular invariants of the $\widehat{SU(3)}$ WZW models at
level $(n-1)$.  Such a relation, if it exists, would not be an absolute
surprise to those who are conversant with the ubiquitous \ADE
classifications of the finite subgroups of $SU(2)$, modular invariants
of the $\widehat{SU(2)}$ WZW, two-dimensional Gorenstein singularities,
and $N=2$ minimal models \cite{CIZ,He-Song,McKay,Ooguri-Vafa}.
 The current situation for $SU(3)$, however, is not nearly as
good as that for $SU(2)$.   That is, unlike the case of $SU(2)$,
 where the \ADE 
Dynkin diagrams precisely classify both the modular invariants and the
finite subgroups, the graphs characterizing the $\widehat{SU(3)}$ modular
invariants are not quite the same as those encoding the irreducible
representations of the finite
subgroups of $SU(3)$; the former graphs appear to be subgraphs of the
latter with many lines and nodes deleted.    
In \cite{DiFrancesco}, it has thus been  suggested
that there should be a way of truncating the representation theory
graphs of the $SU(3)$ 
finite subgroups in order
to reproduce the $\widehat{SU(3)}$ modular invariant graphs, but no
particular algorithm has yet been put forth satisfactorily.

In this paper, we provide evidences for a slightly different
 correspondence using geometry as our main tool.  
The essential motivation for our study stems from
 the fact that  a lot of information about a given finite subgroup
 $\Gamma\subset SU(N)$ are
encoded in the geometry of $\C^N/\Gamma$ and its
 resolution\footnote{Throughout the paper, we use the standard
 mathematical notions of resolution, deformation, and their
 combination as  means of desingularizations.}
 $\pi: M\ra \C^N/\Gamma$.  In particular, string
 theory predicts the Hodge numbers and the Euler characteristics
 of the resolution\footnote{Assuming that
 the crepant resolutions actually exist, that is.}  of such
 Gorenstein orbifolds in terms of group theoretic
 data.  It has subsequently led mathematicians to 
conjecture that there exists a  so-called McKay's 
correspondence between the irreducible representation ring 
 of a finite
 subgroup $\Gamma\subset SU(N)$ and the cohomology ring of the resolved
 manifold  $M$, with certain maps between two ring structures.
Thus, Gorenstein singularities are intrinsically connected to finite
 subgroups of $SU(N)$.  One of the merits of string theory  is that 
(super)-algebra and geometry often play
complementary roles both in its quantization and compactifications.
More precisely, certain non-linear sigma-models on Calabi-Yau (CY) manifolds
 admit purely algebraic descriptions in terms of $N=2$ superconformal
 field  theories.   Therefore, since Gorenstein orbifolds are intrinsically
 related to finite groups, we could perhaps understand the vague connection
 between the modular invariants of $\widehat{SU(N)}$ WZW and the finite
 subgroups of $SU(N)$ if we could find a superconformal field theory
(SCFT) which both encodes the
 WZW theory and describes certain Gorenstein singularities.
This line of thinking furnishes the main theme in the work of Ooguri and Vafa
 \cite{Ooguri-Vafa} and in the present paper: 
Geometry and its SCFT description can provide a possible explanation
for the mysterious connections among the coset models of
the integrable lattices\footnote{We will not explicitly address the
 integrable lattice models in this paper, but it should be possible to
 relate our discussion to those cases which are, in many aspects, 
non-supersymmetric
 cousins of the Kazama-Suzuki theories.   We note that
 both the Kazama-Suzuki models and the continuum limit of the integrable
 lattice models based on $\widehat{SU(3)}$ diagonal modular invariants
 can serve as minimal matters for the (super) $W_3$-algebra.
 Furthermore, both of them are related to Toda and affine Toda
 theories under relevant perturbations.}, 
the WZW modular invariants and the finite
subgroups.

We will restrict ourselves to the case in which the relevant
group is  $SU(3)$ and the SCFT
is a Landau-Ginzburg formulation of the $N=2$ Kazama-Suzuki models.
We will address two complementary questions:  ``Given a
three-dimensional Gorenstein 
singularity, is there a LG theory which describes the non-linear
sigma-model on this geometry and also encodes the $\widehat{SU(3)}$
modular invariants in some way?'' and ``Given a LG theory, is there a
corresponding geometry?'' 
At first sight, it appears that the singularities must be isolated in
order to be related to the well-known Landau-Ginzburg formalism, and
we are consequently led to search for all isolated Gorenstein
singularities. 
We find that the requirement of isolated singularity
imposes a strong constraint on the possible types of three-dimensional
quotient singularities.  It turns out that isolated Gorenstein
singularities in three-dimensions cannot be realized as a hypersurface
in $\C^4$ but only as complete intersections in higher dimensions,
implying that they do not correspond to simple LG theories of the known
type.  
In the opposite direction of pursuit, we face a similar problem; since
LG superpotentials have isolated singularities, there is no direct way
of relating the LG theory to a finite subgroup and its associated orbifold.
We thus take an indirect approach to the problem of relating, if
it is really possible, the LG theory to finite subgroups of $SU(3)$.
What we are able to do is to look for the defining signatures of the
LG theory in the resolved geometry of various Gorenstein orbifolds; in
particular, we look for the chiral ring structure in the 
cohomology of the resolved manifold.  As a result,
we observe some peculiar
matchings between the $SU(3)_n/SU(2)\times U(1)$ Kazama-Suzuki models at
level $n$ and the resolutions of
$\C^3/\Z_{n+3}\times\Z_{n+3}$ Gorenstein orbifolds
which actually have non-isolated
singularities.  Note that the level is different from the conjecture
in \cite{DiFrancesco}.
We also observe that the toric resolution of the
$\C^3/\Z_{n+3}\times\Z_{n+3}$
Gorenstein orbifold yields a natural connection of the
$\Z_{n+3}\times\Z_{n+3}$  finite subgroups
to the modular invariants.  For example, the compact exceptional
divisors, which are degree 6 del Pezzo surfaces\footnote{By a degree 6
del Pezzo surface, we mean $\P^2$ blown up at three points.}, are in
one-to-one correspondence with the chiral primary operators of the
Kazama-Suzuki model, and their intersections reproduce the  
Verlinde algebra and the associated diagonal modular invariant graphs of
$\widehat{SU(3)}_n$ WZW at level $n$.  There are also non-compact
exceptional divisors which are ruled surfaces and which seem to
classify the diagonal $\widehat{SU(2)}_{n+1}$ modular invariants.  We
give an explanation of these observations by noting that
$\Z_{n+3}\times\Z_{n+3}$ has three $\Z_{n+3}\subset SU(2)\subset
SU(3)$ subgroups whose quotient singularities are resolved by the
ruled surfaces.  Thus, our approach, if correct, seems to give us a
canonical way of identifying the relevant elements of the
$\Z_{n+3}\times\Z_{n+3}$ subgroups with the $\widehat{SU(3)}$ modular
invariants.  Namely, those non-trivial 
elements not contained in a subgroup of
$SU(2)\subset SU(3)$ are in one-to-one correspondence with the del
Pezzo surfaces and two-cycles in the resolved manifold, and they are
the ones that classify the $\widehat{SU(3)}_n$ modular invariants.
It thus
seems that there is a mysterious connection between $N=2$ SCFT and
generally non-isolated 
three-dimensional Gorenstein singularities yet to be made more
precise.

This paper is organized as follows:  We begin by discussing what kind
of  subgroups $\Gamma\subset SU(3)$ leads to an isolated singularity
$\C^3/\Gamma$ and the embedding of the resulting geometry as affine
varieties in $\C^n$.  We argue that none of these cases leads to known
LG theories.   \S\ref{sec:KS} initiates the opposite view point and
discusses the relevant details of the $\widehat{SU(3)}$ WZW and
the $SU(3)_n/SU(2)\times U(1)$ Kazama-Suzuki model.  The following
section \S\ref{sec:cznzn} is a search for a corresponding geometry of
the Kazama-Suzuki model.  We analyze the blow-up geometry in both
toric and algebraic geometry set-ups and show that the diagonal modular
invariant graphs appear in the resolution diagrams.   The paper
concludes by addressing some open questions and puzzles.  In Appendix,
we prove a few facts regarding the three-dimensional 
isolated Gorenstein singularities.

\vspace{5mm}

\noindent
{\bf NOTATIONS:}

\noindent
We will adhere to the following conventions throughout this paper: We
define the action of
$\omega := \frac{1}{n}(\alpha_1, \alpha_2, \alpha_3)\in \Z_n$  on
$\C^3$ by $(z_1, z_2, z_3) \mapsto (q^{\alpha_1} z_1, q^{\alpha_2}
z_2,q^{\alpha_3} z_3)$ where $q= \exp (2\pi i/n)$.   
By $\langle\frac{1}{n}(\alpha_1, 
\alpha_2, \alpha_3) \rangle$, we mean a cyclic group of order $n$ generated by
the element $\frac{1}{n}(\alpha_1, \alpha_2, \alpha_3)$.  The
subscripts in $SU(3)_n/SU(2)\times U(1)$ and $\widehat{SU(N)}_n$
represent the levels of the Kac-Moody algebras.   Finally, by
``dimension'', we always mean complex dimension.

%%%%%%%%%%%%%%%%%%%%%%%%%%%%#######################################
%   	Quotient singularities and LG
%%%%%%%%%%%%%%%%%%%%%%%%%%%%#######################################
\newpage
\section{Sigma-Models and Landau-Ginzburg Theories} \label{sec:one}
\setcounter{equation}{0}

%%%%%%%%%%%%%%%%%%%%%%%%%%%%
%
%%%%%%%%%%%%%%%%%%%%%%%%%%%%

\subsection{Isolated Singularities and LG Superpotentials}

It is by now a familiar concept\footnote{One of the first examples
that have been studied describes the quintic in $\P^4$ as a tensor
product of $A_4$ minimal models and plays an important role in 
mirror symmetry.} that certain Calabi-Yau (CY)
sigma-models admit equivalent descriptions in terms of exactly
solvable $N=2$ superconformal field theories (SCFT).  Traditionally,
such an equivalence has usually involved compact projective CY manifolds
and $N=2$ \ADE minimal models.  In \cite{Ooguri-Vafa}, the consideration
has been extended to ALE spaces which are
non-compact two-dimensional CY manifolds.  The natural question to ask
then is whether one can extend the situation to three-dimensional
non-compact CY manifolds obtained from desingularizations of Gorenstein
singularities\footnote{For an elementary introduction to these
terminologies, we refer the reader to \cite{He-Song}.}.  

Let us first briefly review the essential ideas of \cite{Ooguri-Vafa}
and see how to generalize them.  The main ingredient is the fact that
all ALE spaces can be represented as hypersurfaces in $\C^3$, and one
can use the defining equations of those hypersurfaces as
superpotentials in Landau-Ginzburg (LG) theories.  For example, the
singular limit of the $A_n$
ALE space, which is isomorphic to $\C^2/\Z_{n+1}$, can be represented
as $x^2+y^2 +z^{n+1}=0$.  The corresponding superpotential is then
\vspace{1mm}
	\beq
	W = \mu w^{-n-1} + x^2 + y^2 +z^{n+1}  \label{eq:W}
	\eeq
\vspace{1mm}\noindent
where $\mu$ is a moduli parameter and
 the power of $w$ has been chosen\footnote{Recall that the
central charge of a Landau-Ginzburg theory with a superpotential
$W(x_1,\ldots,x_k)$ is given by $c= 3 \sum^k_{i=1} (1-2q_i)$, where
$q_i$ is the $U(1)$ charge of the field $x_i$.  As usual, we define
$3 \hat{c} = c.$}
to yield  the correct total central charge
$\hat{c} =2$ of the sigma-model on the ALE space.  Treating $w$ as a
parameter, we see that \eqr{eq:W} describes a deformation of the $A_n$
singularity.  The claim is that in the degenerating limit, $\mu\ra 0$,
the above SCFT captures the physics of string theory on a singular $ALE$
space.

In a similar spirit, we now search for a possible Gorenstein
orbifold in 3-dimensions whose defining equation can be used as a part
of the LG superpotential.   More precisely, we look for a
desingularization of some quotient
space of the form $\C^3/\Gamma$, where $\Gamma$ is a discrete subgroup
of $SU(3)$, 
whose certain degenerating limit admits a description in terms of a tensor
product of $N=2$ Kazama-Suzuki models \cite{Kazama-Suzuki}.   One of
the criteria for the superpotentials appearing in Landau-Ginzburg theories
is that they must have only isolated singularities\footnote{Having
non-isolated singularities can lead to problems such as an infinite
dimensional chiral ring.}.  Thus, if we insist upon a naive
generalization of the
correspondence between the superpotential and the hypersurface
equation,
we want to consider only those orbifolds with isolated Gorenstein
singularities.   Incidentally in three-dimensions, it turns out that
the requirement of isolated singularity imposes a very restrictive
constraint on the possible types of orbifolds.  Indeed in \cite{Yau-Yu},
Yau and Yu prove the following theorem:

\begin{theorem}\label{theorem:Yau-Yu} 
The three-dimensional Gorenstein singularities of
$\C^3/\Gamma$ are isolated if and only if $\Gamma$ is abelian
and every non-trivial element $g\in\Gamma$ does
not have $1$ as an eigenvalue. \label{theorem:yau}
\end{theorem}

From this theorem, we can determine more precisely 
which  three-dimensional orbifolds of the form $\C^3/\Gamma, \Gamma\subset
SU(3)$,  have only isolated singularities.  Noting that any
abelian finite group $\Gamma$ can be written as a product of cyclic groups,
i.e. \[\Gamma = \Z_{k_1} \times \Z_{k_2} \times \cdots \times
\Z_{k_n}\ , \] we summarize the results, which are proven in the
Appendix, as follows:

\begin{corollary} \label{cor:3}
	The Gorenstein orbifold $\C^3/\Gamma$
 has only isolated singularities if and only if $\Gamma =
\Z_{k}=\langle\frac{1}{k}(\alpha_1,
\alpha_2, \alpha_3)\rangle$ such that GCD$(k,\alpha_i)=1, \forall i$.
This in particular implies that $k$ has to be odd.
\end{corollary}

We would now like to study a possible connection of these Gorenstein
orbifolds with isolated singularities to $N=2$ Landau-Ginzburg
theories.

%%%%%%%%%%%%%%%%%%%%%%
%	Complete Intersections
%%%%%%%%%%%%%%%%%%%%%%

\subsection{$\C^3/\Gamma$ as Hypersurfaces and Complete Intersections}
In order to make a direct connection to Landau-Ginzburg models, as
previously discussed for two-dimensions, we need to determine whether
the Gorenstein orbifolds can be represented by hypersurfaces in
$\C^4$ so that their equations can
play the role of LG superpotentials.  We will now
show that an isolated 3-dimensional Gorenstein singularity actually
cannot be embedded as a hypersurface in $\C^4$ but rather only as a
complete intersection affine variety in higher dimensions.

Let us briefly describe how to embed a Gorenstein orbifold as an affine
algebraic variety in some $\C^n$.  We will closely follow
\cite{Yau-Yu}, restricting our attention to three-dimensions.  
Let $S=\C[z_1,z_2,z_3]$ be a polynomial ring and
$S^{\Gamma}$ its subring of polynomials which are 
invariant under the action of
$\Gamma: (z_1,z_2,z_3)\mapsto \Gamma (z_1,z_2,z_3)$.  One first 
needs to find the minimal set $\{ f_1, f_2, \ldots, f_n\}$ of
generators of $S^{\Gamma}$ as a $\C$-algebra.  Let $\C [y_1,
\ldots,y_n]$ be a  polynomial ring associated to the generators,
then there exists a ring homomorphism 
	\beq
	\upsilon : \C [y_1, \ldots,y_n] \lra S
	\eeq
defined by the substitution map
	\beq
	\upsilon ( F(y_1, \ldots,y_n)) = F(f_1,\ldots, f_n)
	\eeq
where $ F(y_1, \ldots,y_n)\in \C [y_1, \ldots,y_n]$.  Then, we have
$\mbox{Im}(\upsilon) = S^{\Gamma}\cong \C [y_1, \ldots,y_n]/K$, where
$K:= \ker(\upsilon)$ is an ideal with a minimal set of generators
${\cal R}_i$ called
{\it relations}.  Now, the relations define an affine algebraic
subvariety $V_{\Gamma}\subset \C^n$: 
	\beq
	V_{\Gamma} = \{ (y_1, \ldots,y_n)\in \C^n |\ {\cal R}_i(y_1,
	\ldots,y_n)=0, \forall i\} \ .
	\eeq	
It has been shown in \cite{Cartan} that there exists a biholomorphism
$\phi: \C^3/\Gamma \ra V_{\Gamma}$, yielding the desired embedding of
Gorenstein orbifolds as complete intersections in $\C^n$.

Now, using the above prescription, we have explicitly checked 
for many of the isolated Gorenstein singularities given in
Corollary~\ref{cor:3} that they cannot be embedded simply as a
hypersurface in $\C^4$.  For example, we find that 
$\C^3/\Z_3$ embeds in $\C^{10}$.  The computations are very
laborious and discouraging.  
In fact, it turns out that 
isolated singularities of hypersurfaces in four or higher dimensions 
can never be quotient singularities\footnote{This may be related to
Schlessinger's Rigidity Theorem which states that quotient
singularities of codimension three or greater does not have
non-trivial deformations.}  \cite{Yau}.

Thus, we
encounter a difficult problem that the isolated Gorenstein
singularities cannot be simply represented as hypersurfaces in
$\C^4$, and thus, there is no obvious Landau-Ginzburg description of
the singularities.  At best, we can realize them as complete intersections in
higher dimensional embedding spaces $\C^n, n>4$.   It is still
possible to study complete intersections in the LG approach, but it
usually requires many extra variables and a non-trivial change of
variables in the path integral. 
A more serious problem arises from the fact that 
many of the subgroups of $SU(3)$ seem to be ``missing'' in
this analysis in the sense that they have non-isolated singularities
and therefore cannot be related to the
$\widehat{SU(3)}$ WZW modular invariants in this way.  In particular,
we have seen that $\Z_k$, for $k$ even, does not give rise to isolated
singularities and hence cannot be related to a LG superpotential in a
direct way.

%%%%%%%%%%%%%%%%%%%%%%%%%%%%###########################################
%	Modular Invariants
%%%%%%%%%%%%%%%%%%%%%%%%%%%%###########################################
\setcounter{equation}{0}
\section{Modular Invariants and Kazama-Suzuki Models} \label{sec:KS}

It is well-known that the modular invariants of the $\widehat{SU(2)}$
WZW theories fall under an \ADE classification \cite{CIZ}, which
also governs the classification of the finite subgroups of $SU(2)$
\cite{McKay}.  The connection between the
 two {\it a priori}\/ unrelated classifications has been explained in
\cite{Ooguri-Vafa} by using\footnote{up to $U(1)$
projections.} $\frac{SL(2)}{U(1)}\times 
\frac{SU(2)}{U(1)}$ Kazama-Suzuki models
to describe\footnote{Roughly speaking, the $\frac{SL(2)}{U(1)}$ factor
describes the Feigin-Fuchs boson emanating from the singularity while
the  $\frac{SU(2)}{U(1)}$ factor describes the transverse directions
surrounding the singularity.
Furthermore, the $\frac{SU(2)}{U(1)}$ Kazama-Suzuki model is
closely related to the $\widehat{SU(2)}$ WZW theory, and indeed, they
arise from almost identical parafermionic representations.  It is thus
not very surprising that the partition functions of the
$\frac{SU(2)}{U(1)}$ Kazama-Suzuki theory naturally contains the
$\widehat{SU(2)}$ WZW modular invariants, up to different contributions
from the $U(1)$ theta functions.}
 a certain degenerating limit of the orbifolds $\C^2/\Gamma$,
$\Gamma\subset SU(2)$.  
More precisely, the sigma-model/Kazama-Suzuki correspondence states
that the subgroup $\Gamma$  is specified by the same
\ADE Dynkin diagram which classifies the $\widehat{SU(2)}$
modular invariant appearing in the partition function of the
Kazama-Suzuki model.  For example, the $A_{n-1}$ diagonal modular
invariant arises in
the Kazama-Suzuki model at level $(n-2)$ for the $\frac{SU(2)}{U(1)}$ sector, and
the same Dynkin diagram classifies the $\Z_n$ subgroup.  It has been
argued in \cite{Ooguri-Vafa} that the tensored Kazama-Suzuki model at
this level
captures the physics of the orbifold $\C^2/\Z_{n}$ when the $B$-field
has been turned off.  We will consider here the simplest
generalization\footnote{Generalizations to higher dimensions will
appear in \cite{Vafa}.} of \cite{Ooguri-Vafa}, namely
$\frac{SL(2)}{U(1)}\times  
\frac{SU(3)}{SU(2)\times U(1)}$, and see whether there exists a
corresponding geometry of some quotient singularities.  
In this section, we briefly review some useful facts about
the Kazama-Suzuki models, and we will devote the next section to
finding the candidate geometry.  As previously mentioned, 
the hypersurface with an isolated singularity defined by the
superpotential cannot be a quotient singularity in three-dimensions.
Thus, if the Kazama-Suzuki model is related to quotient singularities
at all, then
matching  
the equations of the affine subvarieties representing Gorenstein orbifolds
with superpotentials, as done in \cite{Ooguri-Vafa} for the ALE spaces,
 does not work here, and we need to consider more indirect paths.

%%%%%%%%%%%%%%%%%%%%%%%%%%%%
%   Kazama-Suzuki
%%%%%%%%%%%%%%%%%%%%%%%%%%%%
\subsection{$\widehat{SU(3)}$ WZW and Kazama-Suzuki Models}
\label{subsec:SU(3)-KS} 
The simplest $N=2$ coset models are based on hermitian symmetric
spaces, more specifically complex Grassman manifolds
$SU(n+m)/SU(n)\times SU(m)\times U(1)$.  Algebraically, these types of
Kazama-Suzuki (KS) models are based on the GKO coset construction of
	\beq
	G(k,m,n) = \frac{SU(k+m)_n \times SO(2km)_1}{SU(k)_{m+n}
\times SU(m)_{k+n} \times U(1)_{km(k+m)(k+m+n)}}	
	\eeq
where $n$ is the level of $SU(k+m)$ and so on \cite{Kazama-Suzuki}.  These
models are manifestly 
symmetric in $k$ and $m$, and they actually turn out to be  also
symmetric in any permutation of $k,m,$ and $n$, generalizing the
level-rank duality of WZW models.  The $\widehat{SU(3)}_n$ WZW theory is
related\footnote{In fact, in many ways, G(k,1,n) Kazama-Suzuki models
are $N=2$ generalizations of the $\widehat{SU(k+1)}_n$ WZW theories.}
 to the $G(2,1,n) = SU(3)_n \times SO(4)_1/SU(2)_{n+1}\times
U(1)_{6(n+3)}$ model \cite{Gepner}, which is the one that we will
consider in this 
paper and
denote henceforth by $SU(3)_n/SU(2)\times
U(1)$.  
Before analyzing its Landau-Ginzburg formulation, 
 we will briefly discuss the states and partition functions of
the  model in a more general context.

Let $\Lambda$ be a highest
 weight of $SU(3)$ at level $n$, $a$ that of $SO(4)$ at level 1, and
 $\lambda$  that of $SU(2)\times U(1)$ at level $n+1$.  Then, a
 general field $\Phi^{\Lambda,a}_{\lambda}$ of the coset theory is
 defined by the decomposition
	\beq
	G^{\Lambda} \, V^a = \sum_{\lambda} \Phi^{\Lambda,a}_{\lambda}
 	\, H^{\lambda} \ , \label{eq:decomp}
	\eeq
where $G,V$ and $H$ are fields in the indicated representations of
 $SU(3), SO(4)$ and $SU(2)\times U(1)$, respectively.  The affine
 characters decompose in a similar way, and the character of the coset
 theory is the branching function $\chi^{\Lambda,a}_{\lambda}$ in
	\beq
	\chi^{\Lambda} \, \chi^a = \sum_{\lambda}
 	\chi^{\Lambda,a}_{\lambda} \, 
	\chi^{\lambda} \ .
	\eeq

The modular invariant partition function of the coset theory is then
obtained by taking products of left- and right-handed sectors
	\beq
	Z \ = \ \frac{1}{K}\sum_{\begin{array}{c}
\mbox{\tiny  $\Lambda, \bar{\Lambda}, a, \lambda, \bar{\lambda}$}\vspace{-3mm} \\
\mbox{\tiny  $C(\Lambda, \lambda), C(\bar{\Lambda}, \bar{\lambda})$}\end{array}}
	\chi^{\Lambda,a}_{\lambda}\ {\cal N}_{\Lambda, \bar{\Lambda}}\,
	{\cal M}_{\lambda, \bar{\lambda}} 
	 \ \chi^{\bar{\Lambda},a}_{\bar{\lambda}} ,
	\eeq
where ${\cal N}$ and ${\cal M}$ are  matrices defining the modular
invariants of $\widehat{SU(3)}_n$ and $\widehat{SU(2)}_{n+1}$
WZW, the summation is restricted to satisfy a certain condition
$C(\Lambda, \lambda)$, and $K$ is the order of the
proper external automorphism group\footnote{The factor of $1/K$ is
 included to take care
of the so-called field identification problem.  For the
$SU(m)_n/SU(m-1)_{n+1}\times U(1)$ theory, $K=m(m-1)$.} which identifies fields
in the coset \cite{Gepner2}.

The situation simplifies if we consider only the chiral\footnote{As
usual, a field is chiral if its
 $N=2$ $U(1)$ charge is twice its conformal dimension.} scalars.
Then, the restriction $C(\Lambda, \lambda)$ just requires that
$\Lambda = \lambda$ such that picking the $SU(3)$ integrable highest
weight, after the field identifications, 
uniquely  fixes the weights of $SU(2)\times U(1)$ in the
decomposition \eqr{eq:decomp}.  At level $n$, there are $(n+1)(n+2)/2$
such chiral scalars corresponding to the integrable highest weights,
or primary fields, of the
$\widehat{SU(3)}$ WZW, and they are precisely the scalar components of
the $N=2$ LG superfields realizing the $SU(3)_n/SU(2)\times
U(1)$ KS theory.  We should also note that the LG formulation
corresponds to only the diagonal 
modular invariants, i.e. to diagonal 
matrices ${\cal N}, {\cal M}$.  For non-diagonal modular invariants,
there are generally no corresponding LG descriptions.  Hence, the LG
theory to 
be discussed below captures  the diagonal modular invariants of the
$\widehat{SU(3)}$ WZW theory.

\begin{figure}
\centerline{\psfig{figure=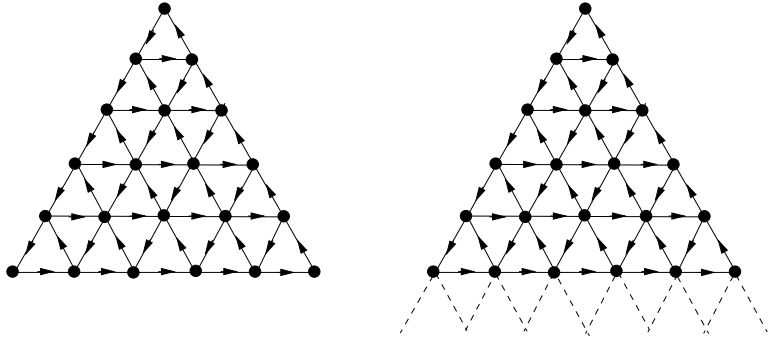,width=4in}}
\caption{ The diagonal $\widehat{SU(3)}_5$ WZW {$\cal A$}-modular invariant
graph at level 5.  For general level $n$, the diagram extends until
there are $n+1$ nodes on each side of the boundary.  In these
diagrams, the nodes represent the primary operators.
\label{fig:modular}}
\end{figure}

The WZW models comprise a very vast and rich subject, and for
the relevant facts regarding the modular invariants and the Verlinde
algebras 
of the $\widehat{SU(N)}$ WZW theories, we refer the reader to
\cite{CFT,He-Song}.  
As an illustration, Figure~\ref{fig:modular} shows an
$\widehat{SU(3)}$ ${\cal A}$-modular invariant at level 5.  These
diagrams will appear again in the toric resolution of $\C^3/\Z_8\times
\Z_8$.

%%%%%%%%
%
%%%%%%%%
\subsection{$SU(3)/SU(2)\times U(1)$ Kazama-Suzuki Model}

Not all Kazama-Suzuki theories have Landau-Ginzburg
realizations, but as mentioned in the previous subsection, 
the particular model of our interest does have one
\cite{Gepner2,LVW}.  Landau-Ginzburg theories capture the chiral
aspects of $N=2$ SCFT, and in the case of $SU(3)_n/SU(2)\times
U(1)$, it is the chiral part that is closely related to the
$\widehat{SU(3)}_n$ WZW theory\footnote{It has been shown in \cite{Gepner}
that generalized Chebyshev integrable deformations of the 
$SU(N)_n/SU(N-1)\times U(1)$ LG theories 
yield the correct Verlinde algebra of the
$\widehat{SU(N)}_n$ WZW.}.  Furthermore, LG theories are known to
describe a different phase of
 CY non-linear sigma-models \cite{Witten}. 
Thus, the $N=2$ LG formulation of the
$SU(3)_n/SU(2)\times U(1)$ KS model is the most natural setting
for studying the connection among finite subgroups, WZW theories, and
geometry.

The superpotentials $W_n$ for the $SU(3)/SU(2)\times U(1)$ Kazama-Suzuki
coset models at level $n$ are given by
	\beq
	W_n (x,y) = q^{n+3}_1 + q^{n+3}_2\ , \label{eq:superpotential}
	\eeq
where it is understood that the superpotential is actually a function
of the symmetric polynomials $x=q_1 + q_2 $ and $y=q_1 q_2$.  In terms
of the $x,y$ variables, the expressions for a few low $n$ are 
\barray
        W_1 &=& x^4  - 4 x^2  y + 2 y^2\nonumber \\
        W_2 &=& x^5  - 5 x^3  y + 5 x y^2\nonumber\\
        W_3 &=& x^6  - 6 x^4  y + 9 x^2  y^2  - 2 y^3\nonumber\\
        W_4 &=& x^7  - 7 x^5  y + 14 x^3  y^2  - 7 x y^3 \nonumber\\
        W_5 &=& x^8 - 8 x^6  y + 20 x^4  y^2  - 16x^2 y^3  + 2 y^4\nonumber\\
        W_6 &=& x^9  - 9 x^7 y + 27 x^5 y^2  - 30 x^3 y^3 + 9 x y^4\nonumber\\
        W_7 &=& x^{10}  - 10 x^8  y + 35 x^6 y^2 - 50x^4 y^3+ 25x^2y^4
                -2y^5 \ ,  \label{eq:w_n}
\earray
where we have rescaled by an over-all normalization.
We see that $W_n$ is quasi-homogeneous if we assign $x$ and $y$ of weights 1
and 2, respectively.
The number of chiral primary fields at level $n$ is
$\frac{(n+2)(n+1)}{2}$ which, as explained before,
 matches the number of primary fields of the $\widehat{SU(3)}_n$ WZW
theory.

In the rest of the paper, we will be interested in a possible 
relation between the above $N=2$ LG theory and 
Gorenstein singularities of the type $\C^3/\Z_{n+3}\times\Z_{n+3}$.  
To produce the correct total central charge $\hat{c}=3$,
we need to
tensor the $SU(3)/SU(2)\times U(1)$ KS to an
extra KS model.
Since we are interested in  non-compact orbifolds, 
with the hindsight from the
two-dimensional black holes \cite{Witten2}, we consider the
$SL(2,\R)_k/U(1)$ model whose central charge at level $k$ is 
$1 + 2/(k-2)$.  Since the central charge of
$SU(3)/SU(2)\times U(1)$ at level $n$ is $2n/(n+3)$,
	\beq
	\hat{c}_{\mbox{\tiny Total}} = \hat{c}_{\frac{SU(3)_n}{SU(2)\times U(1)}} 
	+ \hat{c}_{\frac{SL(2)_k}{U(1)}} = 3\ \ \Longrightarrow\ \ k=
	\frac{n+9}{3} \ .
	\eeq
The superpotential for the LG realization of the $SL(2,\R)_k/U(1)$
model is $W_k = t^{2-k}$.  Hence, the total superpotential for the
tensor product  theory $\frac{SL(2,\R)_{(n+9)/3}}{U(1)}\times
\frac{SU(3)_n}{SU(2)\times U(1)}$ is
	\beq
	W_{\mbox{\tiny T}} (t,x,y,z,w)= \mu t^{-\frac{n+3}{3}}+ q^{n+3}_1 +
q^{n+3}_2 +z^2 + w^2\ ,  \label{eq:W_T}
	\eeq
where again the expression should be understood as a function of
$x$ and $y$, and $\mu$ is a moduli parameter which we set equal to
zero in the singular limit.  Note also that we have added
two more variables whose quadratic terms do not affect the chiral
ring.   Now, as in \cite{Ooguri-Vafa}, we will go to the patch where
$t\neq 0$ and think of $W_{\mbox{\tiny T}} (1,x,y,z,w)=0$ as defining
a hypersurface in $\C^4$, which we hope to relate to certain
Gorenstein orbifolds.

%%%%%%%%%%%%%%%%%%%%%%%%%%%%###########################################
%   Z_n x Z_n
%%%%%%%%%%%%%%%%%%%%%%%%%%%%###########################################

\setcounter{equation}{0}
\section{$\C P^2$ Kazama-Suzuki and $\cznzn$} \label{sec:cznzn}
In this section, we present some evidences that the Landau-Ginzburg 
formulation of the
$SU(3)_n/SU(2)\times U(1)$ KS model may be related to a
desingularization of  the 
$\C^3/\Z_{n+3}\times\Z_{n+3}$ orbifolds.  Because of
Theorem~\ref{theorem:Yau-Yu},  
given a KS LG superpotential, or a hypersurface with an isolated
singularity for that matter, 
there is no canonical way of relating it to a finite
subgroup of $SU(3)$ and its associated Gorenstein orbifold.

It is important to note at this point that the natural physical
degrees of freedom in a LG theory are the complex deformations of the
superpotential by chiral ring elements.  The  KS LG theories thus, if
they are related to Gorenstein orbifolds in some way, describe a
deformation, not resolution, of the singularities.  Deformation and
resolution of singular orbifolds in three-dimensions are two very
different processes, generally leading to topologically distinct
manifolds.  Thus, a resolution of an orbifold does not necessarily
carry information about the structure of a deformed manifold; 
but amazingly, it sometimes does.  An example is the
phenomenon occurring in \cite{Vafa-Witten} in the context of 
discrete torsion. In that case, $T^2\times T^2\times T^2/\Z_2\times 
\Z_2$ can be either completely resolved without discrete torsion or 
deformed in terms of the invariant variables in the presence of discrete 
torsion, resulting in 64 remaining conifold singularities.  The two 
desingularizations are argued to be mirror pairs. 
Motivated by this interesting case and the ease with which resolutions
can be studied, we  propose a simple but naive step
towards finding possible subgroups of $SU(3)$ that can be related to
the KS LG theories:  We search for resolutions of $\C^3/\Gamma,
\Gamma\subset SU(3)$, whose classical cohomology encodes the chiral
ring structure of the $SU(3)_n/SU(2)\times U(1)$ Kazama-Suzuki model.
An {\it ad hoc}\/ proposal such as ours is worth considering if and perhaps
only if it meets a success; but surprisingly, it has.
In this section, we will see that the resolution of $\cznzn$  seems to
be closely related to our coset theory.

It turns out that considering the subgroups of the form
$\Z_n\times\Z_n$ cures some of the difficulties previously
encountered, while, at the same time, introducing new obstacles.  The 
group 
	\beq
	\znzn =\langle\, \frac{1}{n+3}(\alpha, 
-\alpha, 0)\ ,\ \frac{1}{n+3}(0,\alpha, 
-\alpha)\, \rangle \ ,
	\eeq
where $\exp(2\alpha\pi i /(n+3))$ is a primitive $(n+3)$-th root of unity,
is actually a maximal finite subgroup of
$SU(3)$ consisting of all cyclic elements of order $(n+3)$.  Thus, we seem
to have incorporated many of the ``missing'' subgroups of $SU(3)$ in
this approach.  On the other hand, the orbifold
$\cznzn$ has singularities along subvarieties, and thus, their
defining hypersurface equations cannot be used
as LG superpotentials at first sight\footnote{Such superpotentials
lead to degenerate conformal field theories, and the chiral ring is
infinite dimensional.}.  
We nevertheless argue that the
resolved geometry of that type appears to classify the
$\widehat{SU(3)}$ as well as $\widehat{SU(2)}$ WZW modular
invariants. 

We first study the geometry of the resolution of $\cznzn$
orbifold singularities both from the toric geometry and algebraic
geometry points of view.  We then comment on the desingularization by
complex structure deformations and speculate that the LG
theories describe the geometry of certain deformations of $\cznzn$. 

%%%%%%%%%%%%%%%%%%%%%%%%%%%%
%
%%%%%%%%%%%%%%%%%%%%%%%%%%%%

\subsection{Cohomology of the Resolution and McKay Correspondence}
There are several ways to study the resolution of three-dimensional
Gorenstein singularities and compute the  Hodge numbers of the
resolution.
It is known that all crepant resolutions are related to
each other by flops.
Let $\pi: M \ra\cznzn$ be a resolution of $\cznzn$.
Then, using the {\it age grading}\/ of the $\znzn$ subgroup
introduced in \cite{Ito-Reid}
in the context of McKay correspondence\footnote{For a review on McKay
correspondence, see \cite{He-Song,Ito-Reid}.},
we find:
	\barray
	\#\{\mbox{elements of age 0}\} &=& h^0(M,\Q)\ =\ 1 \nonumber\\
	\#\{\mbox{elements of age 1}\}  &=& h^2(M,\Q)\ =\
		\frac{(n+7)(n+2)}{2}\nonumber\\
	\#\{\mbox{elements of age 2}\} &=& h^4(M,\Q)\ =\
	\frac{(n+1)(n+2)}{2}\ ,  \label{eq:hodge}
	\earray
where $h^i$ are the Hodge numbers of the resolved manifold.  These
numbers can also be obtained from toric geometry, as will be discussed
in the next subsection.

The sum of the Hodge numbers, which is just the Euler characteristics
of $M$ in this case, is equal to $(n+3)^2$ which is precisely the order of
the group $\znzn$.  The number $(n+3)^2$ is also the number of irreducible
representations of $\znzn$, and thus, we see that 
the predictions of the McKay correspondence--or string theory, whichever
the reader prefers--are well satisfied.

%%%%%%%%%%%%%%%%%%%%%%%%%%%%
%
%%%%%%%%%%%%%%%%%%%%%%%%%%%%
\subsection{Toric Resolution of $\cznzn$}

\begin{figure}[hbt]
\centerline{\psfig{figure=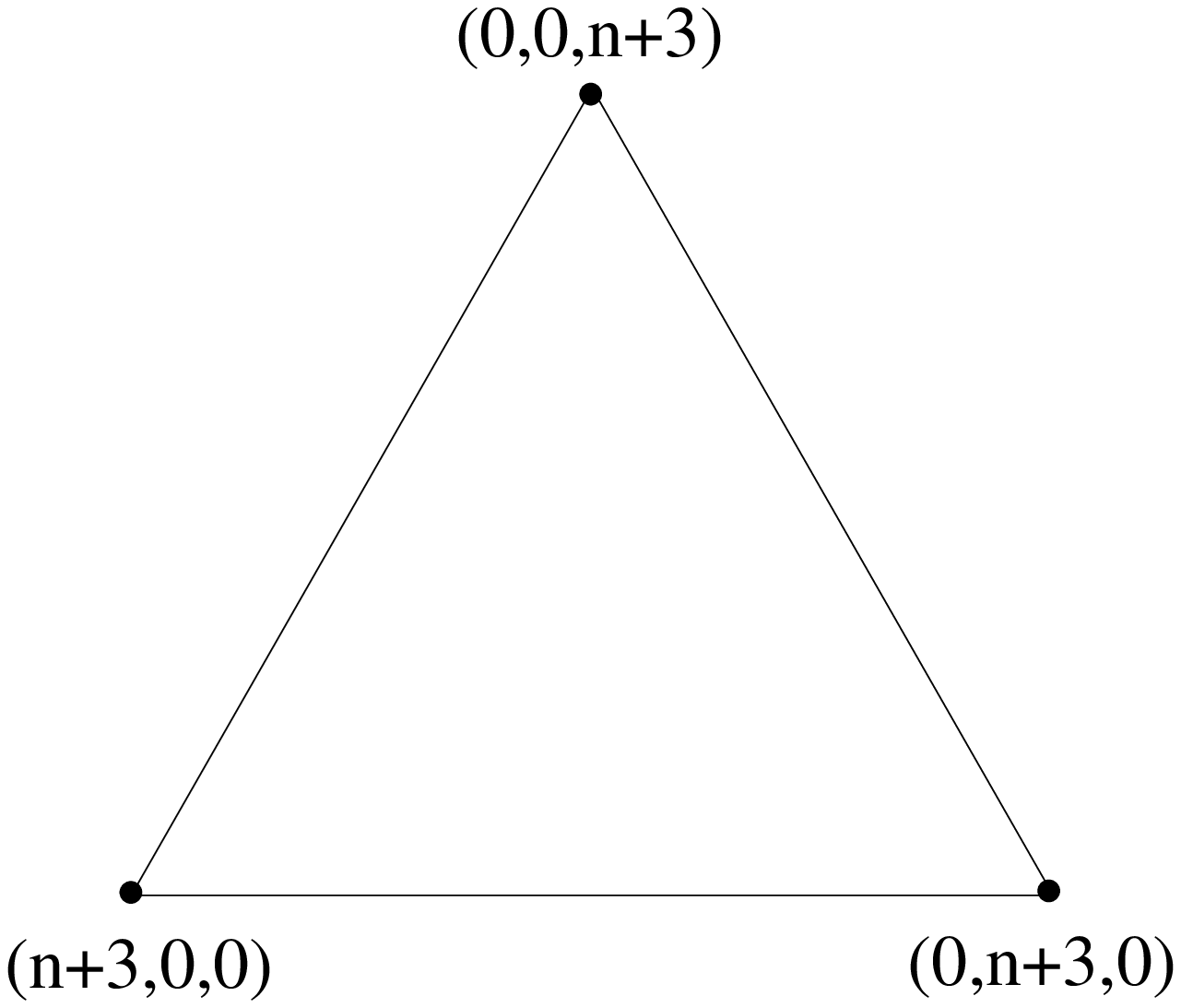,width=2in}}
\caption{The polygon for unresolved $\cznzn$.
\label{fig:fan}}
\end{figure}

The Hodge numbers \eqr{eq:hodge} can also be computed directly from
geometry by using toric blow-ups and Poincar\'{e} duality.  For
background materials on toric varieties, see \cite{Oda}.
The cone for the unresolved $\cznzn$ is defined over a convex polygon
given in Figure~\ref{fig:fan} which is a hyperplane cross-section of
the first ``quadrant'' of
the standard integral lattice $\Z^3$ by a plane passing through
$(n+3,0,0), (0,n+3,0),$ and $(0,0,n+3)$.

\begin{figure}[hbt]
\centerline{\psfig{figure=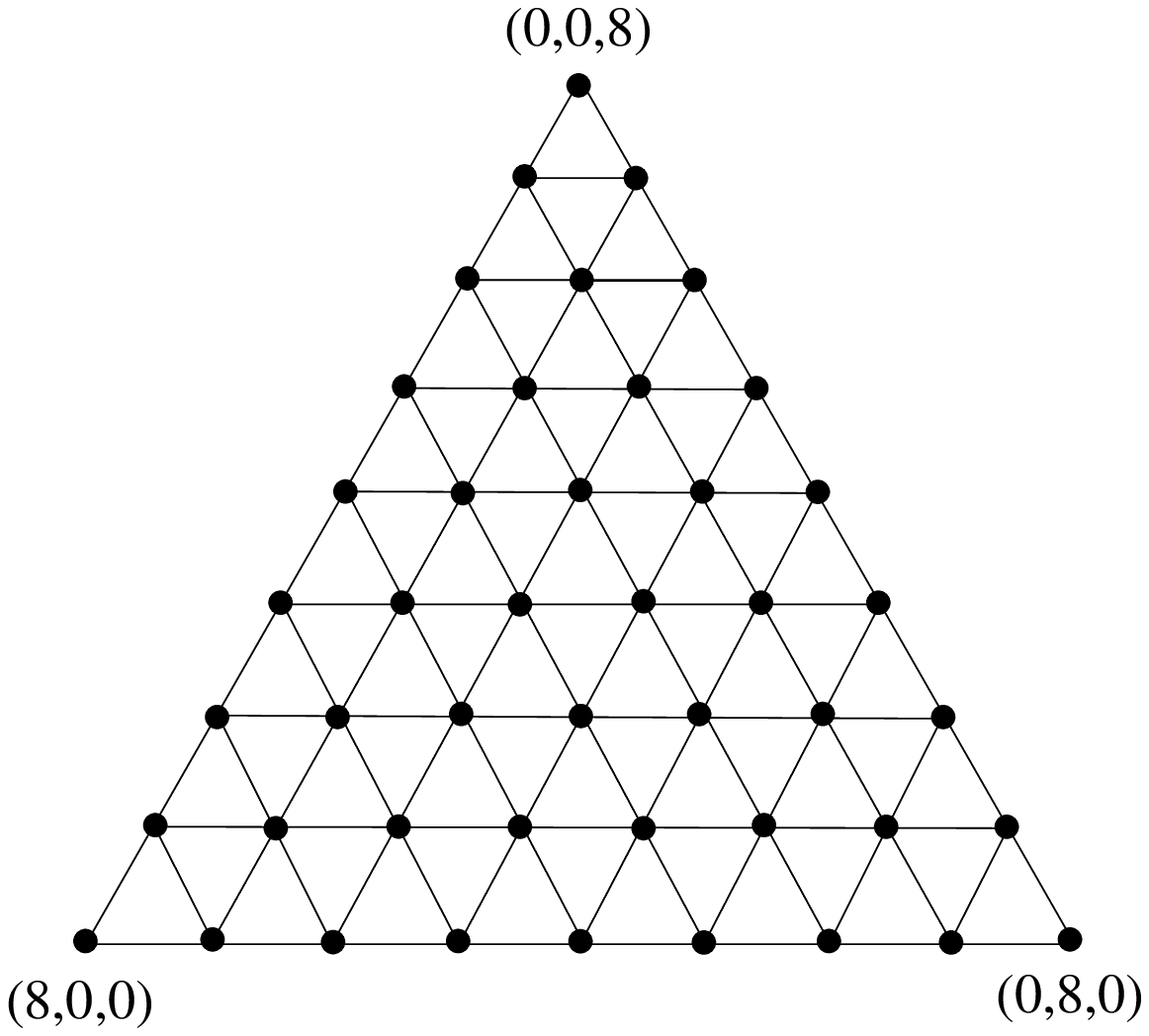,width=2.5in}}
\caption{The polygon for completely resolved $\C^3/\Z_8\times\Z_8$.
The toric fan consists of cones defined over this polygon.
\label{fig:resolved-fan}}
\end{figure}

To resolve the singularities, we need to add in all 
lattice points lying on the polygon  and triangulate the cone to produce
triangles with unit area in appropriate units.  As an example,
Figure~\ref{fig:resolved-fan} gives
the polygon for a particular complete resolution of $\C^3/\Z_8\times \Z_8$,
other crepant resolutions being related to this one by a sequence of 
flops.

Here, the new points on the outer boundary of the polygon represent
non-compact exceptional divisors which are ruled surfaces lying along
the three coordinate axes of $\C^3/\Z_8\times\Z_8$.  In the general
case of $\cznzn$, we will have $3(n+2)$ ruled surfaces as
non-compact exceptional divisors.  We can understand their origin as follows:
The group $\znzn$ has two generators 
$\omega=\frac{1}{n+3}(\alpha, -\alpha, 0)$ and $\eta= \frac{1}{n+3}(0,\alpha, 
-\alpha)$.  Then, there are three $\Z_{n+3}\subset SU(2)$ subgroups of $\znzn$, 
namely $\langle \eta\rangle$,
$\langle \omega\eta\rangle$, and
$\langle \omega \rangle$, which fix $z_1, z_2, $ and $z_3$ coordinates of
$\C^3$, respectively,
and produce the familiar $A_{n+2}$ ALE singularities along those
axes.  The $A_{n+2}$ singularities in ALE spaces are resolved by a
chain of $(n+2)$ $\P^1$ blow-ups which, upon fibration over the
coordinate axes, become the $3(n+2)$ ruled 
surfaces that we see in the toric
resolution.  Now, since $A_{n+2}$ ALE singularities classify the
diagonal $\widehat{SU(2)}_{n+1}$ WZW modular invariants, what we have
just found is that the ruled surfaces in the resolution of $\cznzn$
classify the same invariants.

The  remaining $(n+1)(n+2)$ non-trivial elements of
$\znzn$ not contained in 
those three $\Z_{n+3}\subset SU(2)$ subgroups are all seen not to possess
an eigenvalue
1, and thus, they lead to an isolated quotient singularity at the
origin which is resolved by introducing $(n+1)(n+2)/2$ degree 6 del
Pezzo surfaces.    These compact exceptional divisors are the 21
interior 
points in Figure~\ref{fig:resolved-fan}, which is the correct number of
del Pezzo surfaces for $n=5$.  
Immediate from the toric digram
is another observation that the
$\widehat{SU(3)}_{n}$ 
modular invariants are classified by the $(n+1)(n+2)/2$ 
compact del Pezzo exceptional
divisors.
The fourth Hodge number $h^4$ can be computed by applying the
Poincar\'{e} duality to the compact\footnote{Because we are dealing
with non-compact spaces, we need to take some caution when applying
mathematical facts that are familiar from studying compact spaces.} 
second cohomology $H^2_c(M,\Q)$, which is dual
to the del Pezzo surfaces, and thus we have 
$(n+1)(n+2)/2$
 two-cycles which appear in the intersections of the exceptional divisors.
We now see that of $(n+1)(n+2)$ non-trivial elements of
$\znzn$ not contained in the $\Z_{n+3}\subset SU(2)$ subgroups, half of
them corresponds to the del Pezzo surfaces and the other half to the
non-trivial two-cycles.
Furthermore, the aforementioned McKay correspondence is clearly
satisfied, and there 
is indeed a one-to-one correspondence between the irreducible
representations of  $\znzn$  and the cohomology elements of the
resolved manifold \cite{Ito-Reid}.

%%%%%%%%%%%%%%%%%%%%%%%%%%%%
%
%%%%%%%%%%%%%%%%%%%%%%%%%%%%
\subsection{Intersection Homologies and Blow-ups}
We now study  the classical intersection theory on the
resolved manifold and see that it reproduces a perturbed 
chiral ring structure of the coset theory.  
We do not know 
whether the full quantum intersection theory would correspond to the
unperturbed chiral ring, which happens to be the cohomology ring of the
Grassmannian $U(n+2)/U(2)\times U(n)$ satisfying the Schubert calculus.

In this section, we argue
that the classical intersection
homology of the resolved $\C^3/\Z_{n+3}\times \Z_{n+3}$
 captures the diagonal Verlinde algebras of the
$\widehat{SU(3)}_n$ and $\widehat{SU(2)}_{n+1}$ WZW theories.  
There are two ways of illustrating this.  The first way is to use the
Stanley-Reisner relations in toric geometry to find the cohomology
ring structure.   In fact, without going  into details, 
a rough picture of the intersection
homology comes from the following facts:
	\begin{enumerate}
	\item Two divisors, which are represented by points on the
polygon, intersect along a two-cycle if and only if their points are
connected by a line in the polygon. 
	\item Three divisors intersect at a point if and only if their
corresponding points form three vertices of a triangle in the
triangulation of the polygon.
	\end{enumerate}
Hence, just by looking at the toric resolution diagrams,
 the simple-minded rule that two primary fields fuse if and only
if their corresponding exceptional divisors intersect gives a
relation\footnote{This rule is a generalization of that in
two-dimensions where the intersection matrices of exceptional $\P^1$
divisors reproduce the fusion matrices of $\widehat{SU(2)}$ WZW.  But,
here the analogy is not completely satisfactory, because the divisors
intersect along any one of the  $(n+1)(n+2)/2$ two-cycles in the
resolution of $\cznzn$.} 
between the fusion matrices of the $\widehat{SU(3)}$ and
$\widehat{SU(2)}$  WZW theories and the intersection
homology of the resolved manifold.  That is, the intersections of the
ruled surfaces give the Verlinde algebra of the
$\widehat{SU(2)}_{n+1}$ WZW while the intersections of the del Pezzo
surfaces yield that of the $\widehat{SU(3)}_n$ WZW.

The second way to see the details of the resolved geometry of
$\cznzn$ is to blow up explicitly along subvarieties which are loci of
the singularities.  We will just sketch the main ideas by blowing up
along one of the coordinate axes.
As will be subsequently discussed, the equation 
$xyz=w^{n+3}$ describes the orbifold $\cznzn$ as an affine subvariety
$M\subset \C^4$.  It
 has three lines of
singularities along $xy=0, xz=0$, and $yz =0$.  Without a loss of
generality, let us take the first
locus $xy=0$ and blow up along the $z$-axis.  Define
	\beq
	\Delta = \{ (x,y,z,w) \times (s,t,u) \in \C^4\times \P^2\ |\
xt=ys,\, xu=ws,\, yu=wt\} \ . \label{eq:Delta}
	\eeq
We have effectively replaced $\C^4$ by $\C^1\times {\cal
O}_{\P^2}(-1)$, or equivalently, we have replaced the origin of $\C^4$
with an exceptional divisor $E=\P^2$ so that $\pi: \Delta\setminus
E\ra \C^4 \setminus\{0\}$ is an isomorphism.  Away from $x=y=w=0$,
$\pi^* M$ is just its 1-1 pre-image.  To understand the blow-up
geometry at $x=y=w=0$ and $z\neq 0$, we consider a line
passing through a point $(x,y,z,w)$ on the hypersurface and 
$(0,0,z,0)$ at constant $z$.  By construction
\eqr{eq:Delta}, each such a line defines a point on the $\P^2$.  Now,
we take the limit where the point on the hypersurface approaches
$(0,0,z,0)$ in all possible directions 
and determine the corresponding limiting points on the
$\P^2$.  Intuitively, each tangent line at the singular
point $(0,0,z,0)$ along the constant $z$ subvariety
becomes a point on the exceptional divisor $\P^2$, and thus in the
blow-up picture the
hypersurface will end on a one-dimensional subvariety on $\P^2$ as
follows:   Pick a point $[s,t,u]\in\P^2$ with a fixed length $|s|^2 +
|t|^2 + |u|^2 =1$, 
and consider the point $(\epsilon s, \epsilon t ,z,\epsilon
u)\in\C^4$, where $|\epsilon| << 1$.  Now, demanding that the point
$(\epsilon s, \epsilon t ,z,\epsilon u)$ lies on the hypersurface
$xyz=w^{n+3}$, we have
	\beq
	 stz \epsilon^2 \ = \ \epsilon^{n+3} u^{n+3} \approx 0\ \ \Lra
\ \ s = 0 \ \mbox{ or }\  t=0.
	\eeq
Hence, we see that the hypersurface ends on two intersecting $\P^1$'s
in $\P^2$ defined by the homogeneous coordinates $s=0$ and $t=0$.
Since the singularities actually occur along the $z$-axis, each $\P^1$
is fibrated over the axis and thus defines a ruled surface.

We also want to know how the hypersurface has transformed near the
exceptional $\P^2$, and for that purpose, we need to view $\Delta$ as
 $\C^1\times {\cal O}_{\P^2}(-1)$, where ${\cal O}_{\P^2}(-1)$ is the
universal bundle of $\P^2$.
On the patch where $u\neq 0$, the good
coordinates of the $\P^2$ are	
	\beq
	\alpha = \frac{s}{u} \ \ \mbox{and} \ \  \beta =\frac{t}{u}.
	\eeq
In these variables, we have $x=w\alpha$ and $y=w\beta$, and the
hypersurface becomes
	\beq
	xyz=w^{n+3}\ \ \Longrightarrow \ \ w^2 (\alpha\beta z -w^{n+1}) =0
\ . \label{eq:new-hyper}
	\eeq
In the new coordinates, $w=0$ corresponds to $x=y=w=0$ and thus to the
entire $\P^2$.  So, the intersection of the hypersurface with $\P^2$
is given by
	\beq
	\alpha\beta z -w^{n+1} =0, \label{eq:new-hyper2}
	\eeq
which is just the original equation with its degree diminished by 2,  
and the resulting singularity structures 
\eqr{eq:new-hyper2} are the same as before.   That is, it again has
singularities along the $z$-axis at $\alpha\beta=w=0$, which is
precisely at $s=t=0$ where two $\P^1$'s meet on $\P^2$.  
This kind of
singularities is exactly the same as that appearing in 
the resolution of $A_{n+2}$ ALE spaces, except here we introduce a pair
of ruled surfaces rather than $\P^1$'s with each blow up.  Repeating
this procedure until we have resolved all the singularities along the
$z$-axis will thus produce $(n+2)$ ruled surfaces intersecting in a
chain, which clearly corresponds to the points on the outer edge of
the toric picture.  We have now understood the 3 chains of $(n+2)$
ruled surfaces whose intersections clearly resemble the Verlinde
algebra of the $\widehat{SU(2)}_{n+1}$ WZW theory.  In retrospect, 
we should  have expected this
phenomenon because, as previously discussed, these non-isolated
singularities result from the $\Z_{n+3}\subset SU(2)\subset SU(3)$ and we
know from $\cite{Ooguri-Vafa}$ that $A_{n+2}$ ALE spaces classify the
$\widehat{SU(2)}_{n+1}$ diagonal modular invariants.

Analysis of the extra singularity at the origin can be performed in a
similar way, in principle, but the computation is highly non-trivial
and we omit its presentation in this paper.  For that purpose, it is
much easier to resort to the toric resolutions.  We have previously
seen that the extra quotient singularities at the origin arise from
those elements of $\znzn$ which are not contained in a $SU(2)$
subgroup of $SU(3)$.  Half of those non-trivial elements corresponds
to the $(n+1)(n+2)/2$ del Pezzo surfaces, and the other half should
correspond to $(n+1)(n+2)/2$ two-cycles, which we have not been able to
analyze in any detail\footnote{To do so, one really needs to study the
Stanley-Reisner ideals to find the cohomology generators.}. 
Thus, if we were to relate the finite subgroups
$\znzn$ to the diagonal
$\widehat{SU(3)}_n$ WZW modular invariants, we need to
look at the del Pezzo divisors over the origin.  Indeed, the
intersections of those del Pezzo surfaces seem to reproduce the
Verlinde algebra of $\widehat{SU(3)}_{n}$ WZW at level $n$ and 
classify the corresponding diagonal modular invariant.  A 
more precise formulation of the correspondence would have to realize
the fusion coefficients as some functions\footnote{Our speculation
that the primaries fuse if and only if the del Pezzo surfaces
intersect along some two cycle seems  too naive at the
moment.} of 
the intersection homology, which would require an understanding of the
two-cycles.

\begin{figure}
\centerline{\psfig{figure=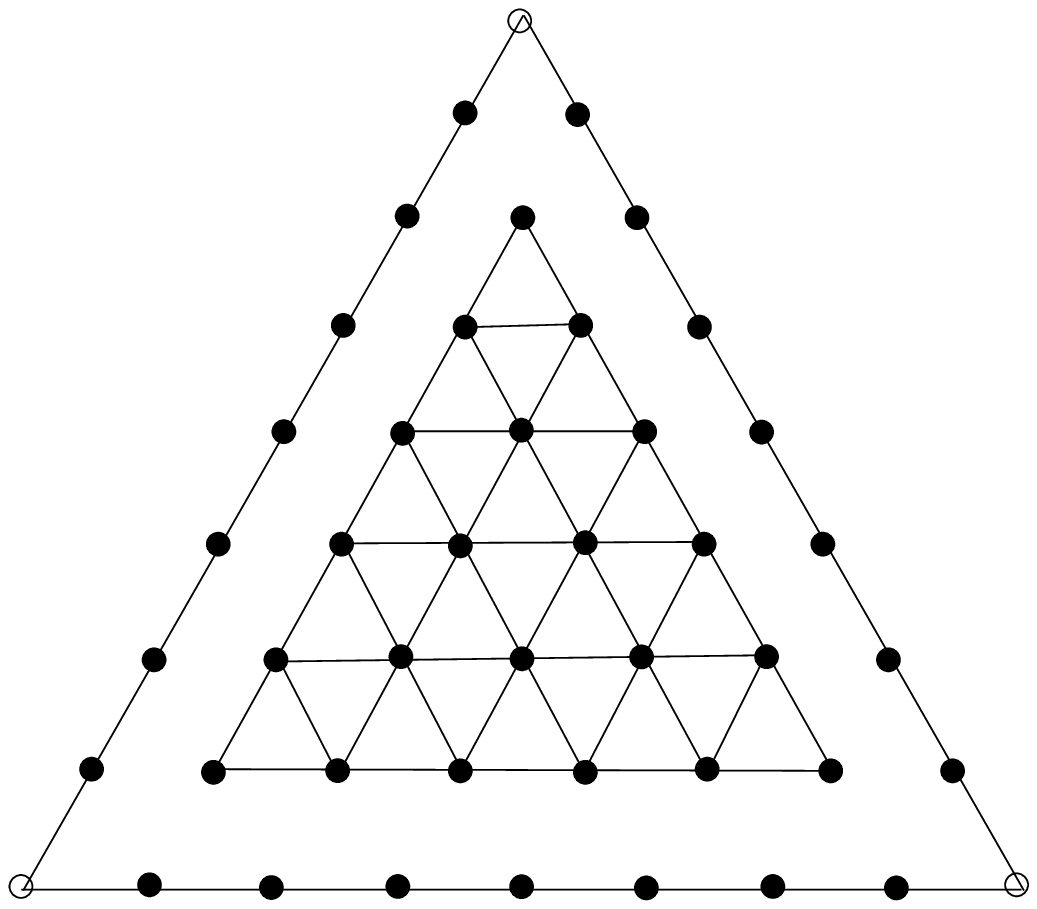,width=3in}}
\caption{The fan for the resolved $\C^3/\Z_8\times\Z_8$ encodes
the $\widehat{SU(3)}_5$ and $\widehat{SU(2)}_6$ modular invariants.
\label{fig:su2su3}}
\end{figure}

We now need to know how the
$\widehat{SU(2)}_{n+1}$ and  $\widehat{SU(3)}_n$ WZW theories are
related to the $SU(3)_n/SU(2)\times U(1)$ Kazama-Suzuki model.
Besides the fact that these theories arise
in the GKO coset construction of the
 KS theory, 
recall the following facts about the WZW models and
the LG formulation of the KS theory \cite{Gepner}:  
The $SU(3)_n/SU(2)\times U(1)$ Kazama-Suzuki model has
$(n+1)(n+2)/2$  chiral primary fields 
and  under the perturbation of the LG
superpotential by generalized Chebyshev polynomials, the chiral ring
structure reproduces the  $\widehat{SU(3)}_n$ WZW 
fusion coefficients.  Hence, the chiral
primaries of the KS theory are in one-to-one correspondence with the
primary fields of the $\widehat{SU(3)}_n$ WZW theory, and the chiral
algebra is the homogeneous part of the Verlinde algebra.  This
correspondence between the deformed chiral ring and the WZW Verlinde
algebra actually generalizes to all $\C P^N$ KS and
$\widehat{SU(N+1)}$ WZW.
Furthermore,
the Landau-Ginzburg formulation of the Kazama-Suzuki theory at level
$n$ is obtained from 
the diagonal modular invariants of the $\widehat{SU(3)}_n$ WZW
at level $n$, which  is classified by diagrams such as
Figure~\ref{fig:modular}, and those of the $\widehat{SU(2)}_{n+1}$ WZW
at level $(n+1)$, which is classified by the $A_{n+2}$  Dynkin diagram.
Incidentally, we have just seen that  the toric resolution diagram for
$\C^3/\Z_{n+3}\times\Z_{n+3}$ contains both the 
$\widehat{SU(3)}_n$ WZW modular invariant graph as a subgraph
describing the del Pezzo surfaces and the $\widehat{SU(2)}_{n+1}$ WZW
modular invariant graphs on the outer edges.  For $n=5$, we
display the observation in Figure~\ref{fig:su2su3}.  Whether this
strange, but general,
phenomenon is a complete fluke or is actually in line with the
attempt to classify the $\widehat{SU(3)}_n$ WZW modular invariants using
finite subgroups remains to be seen.  
In order to prove that there actually exist underlying 
relations among the WZW
modular invariants, the subgroups $\Z_{n+3}\times\Z_{n+3}$ and the
orbifolds $\C^3/\Z_{n+3}\times\Z_{n+3}$, we would need a SCFT
description of the Gorenstein orbifolds.
Turning things around, we  speculate that, given the close
connections between the KS and WZW models, the above observations seem
to suggest that the SCFT is likely to contain the $SU(3)_n/SU(2)\times
U(1)$ Kazama-Suzuki model as one of its factors.  The LG SCFT however
does not describe the resolution of $\cznzn$, but rather some
deformation or a combination of both.

We devote the remainder of the paper to examining 
the above observations and their consequences.

%%%%%%%%%%%%%%%%%%%%%%%%%%%%
%
%%%%%%%%%%%%%%%%%%%%%%%%%%%%

\subsection{Hypersurfaces and LG Theories}

Our ultimate interest in studying any kind of correspondences among
seemingly unrelated objects lies in understanding the {\it a priori}
reason for such occurrences.  The status of the $SU(2)$ cases is much
more well-founded than the fairly untouched $SU(3)$ counterparts.
Thus, without any pretense of rigor or fallacious confidence, we
want to devote the rest of this paper in making several comments
which may shed some light for future efforts.

For our study, the work of Joyce plays an important role \cite{Joyce}.
Given a Gorenstein orbifold $\C^3/\Gamma$, we can often find a family of
desingularizations $\pi: M \ra \C^3/\Gamma$ carrying geometric
structures that are compatible with Calabi-Yau conditions and which
approach the orbifold geometry at a degenerating limit.  Using
deformations of codimension two singularities, Joyce has
shown that there are in fact many topologically distinct families of
Calabi-Yau 
desingularizations, with different\footnote{It is believed that all
crepant resolutions in any dimension give rise to the string theory
orbifold Euler characteristics and Hodge numbers, when things are
properly defined.}
Hodge numbers and Euler characteristics.  Indeed, mathematically,
desingularizing orbifold singularities is an extremely complicated and
laborious process.    Even for the simplest case of
$\C^3/\Z_2\times\Z_2$, Joyce has found nine distinct Calabi-Yau 
desingularizations.  Similarly, in the case of $T^6/\Z_2\times\Z_2$,
there are thousands of different Calabi-Yau desingularizations, two of
which are selected out by string theory with and without a discrete
torsion \cite{Vafa-Witten}.  Why do other desingularizations
not have physical realizations?  Is it possible that there are
physical desingularizations of which we are presently unaware?

So far, we have seen that the resolution of the $\cznzn$ Gorenstein
orbifold seems to reproduce the Verlinde algebras and the modular
invariants of the $\widehat{SU(3)}_n$ and $\widehat{SU(2)}_{n+1}$ WZW
theories, which arise in the construction of the $SU(3)_n/SU(2)\times
U(1)$ KS model.  We interpret this phenomenon as implying
that the KS LG theory could be
describing some kind of a deformation of $\cznzn$ in
such a way that upon blowing down the the exceptional divisors in the
resolution and then deforming the singularities, the information about
the intersection homology on the resolved manifold gets transmitted to the
chiral ring of the KS theory.  It is not clear to us how to
interpret the chiral ring geometrically in terms of the deformed
manifold, although as will be subsequently discussed,
 studying the intersections of vanishing cycles on
the Milnor lattice of the KS LG affine variety gives a suggestion.  
Interestingly, we have found a coordinate transformation which
transforms the KS superpotential into the hypersurface equation for
$\cznzn$, but the transformation is not everywhere well-defined.

The first step in trying to desingularize the Gorenstein singularity
by deformation is to
embed the orbifold in $\C^4$ as a hypersurface, and then, deform
the algebraic equation while maintaining the Calabi-Yau properties.
Using the procedure of \cite{Yau-Yu}, we can represent the orbifold
$\cznzn$ as a hypersurface in $\C^4$ as follows:  The independent,
invariant monomials that can be constructed out of $(z_1,z_2,z_3) \in \C^3$
are 
	\barray
	x &=& z_1^{n+3} \nonumber\\
	y &=& z_2^{n+3} \nonumber\\
	z &=& z_3^{n+3} \nonumber\\
	w &=& z_1 z_2 z_3 \ ,
	\earray
and it can be checked that they generate the invariant subring
$S^{\znzn}$ of the polynomial ring $\C[z_1,z_2,z_3]$.
The relation among the invariants is
	\beq	
	xyz = w^{n+3} \ , \label{eq:xyz=w^n}
	\eeq
which embeds $\cznzn$ in $\C^4$.

Now, note the curious fact that there is a singular coordinate transformation
that maps the Kazama-Suzuki superpotential to an expression similar to the form
\eqr{eq:xyz=w^n}:  Consider the redefinitions
	\beq
	x = \beta^{-\frac{1}{n+3}} (1+\alpha) \ \ , \ \ y = \alpha
	\beta^{-\frac{2}{n+3}}	\label{eq:transformation}
	\eeq
in the superpotentials \eqr{eq:w_n} or  \eqr{eq:superpotential}.  Then, the
superpotentials can be written as
	\beq
	W_n = \frac{1}{\beta}\left( 1+ \alpha^{n+3} + \beta zw \right) \ ,
	\eeq
where the term inside the parenthesis is a deformation of the 
 expression \eqr{eq:xyz=w^n}.  It seems to suggest that the
$SU(3)_n/SU(2)\times U(1)$ Kazama-Suzuki model is indeed describing a complex
deformation of the Gorenstein orbifold $\C^3/\Z_{n+3}\times\Z_{n+3}$,
but we
 do not understand the validity of this
argument since the coordinate transformation has a non-constant Jacobian
	\beq
	\frac{\partial (x,y)}{\partial (\alpha,\beta)} \ =\
\frac{\beta^{-\frac{3}{n+3}-1}}{n+3}\ (1-\alpha)
	\eeq
which is not everywhere well-defined.  

The motivation for the
coordinate transformation \eqr{eq:transformation} comes from the fact
that the  $SU(3)_n/SU(2)\times U(1)$ Kazama-Suzuki  model possesses a
discrete $\Z_{n+3}$ symmetry and thus, as it is familiar from standard Gepner
constructions, that we really need to consider a $\Z_{n+3}$ orbifold of
the KS LG theory.  It is easy to see that the coordinates $\alpha$ and
$\beta$ are $\Z_{n+3}$ invariant coordinates.

Besides the deformed chiral ring structure of the KS theories, there are
other evidences that the KS model is an $N=2$ analogue of
the $\widehat{SU(3)}_n$ WZW theory.
For example, regarding the $SU(3)_n/SU(2)\times U(1)$ 
KS LG superpotential as defining an affine variety
actually leads to very interesting results.  In \cite{GZV}, it has
been shown that the intersection form of vanishing cycles in the
Milnor lattice of the affine variety defined by the superpotential
reproduces the  $\widehat{SU(3)}_n$ Verlinde algebra.
Thus, we see that the classical intersection theory of the del Pezzo
surfaces in the resolved
Gorenstein orbifold is encoded in the intersection forms of vanishing
cycles in the affine variety of the KS LG superpotential.

%%%%%%%%%%%%%%%%%%%%%#############################################
%	Conclusion
%%%%%%%%%%%%%%%%%%%%%#############################################

\section{Conclusion: Much Ado about Nothing?} \label{sec:conclusion}
The idea of classifying the $\widehat{SU(3)}$ WZW modular invariants in
terms of finite subgroups of $SU(3)$ has not met much success to
date.  In this paper, we have reduced the problem into geometry and
$N=2$ superconformal field theory--that is, into finding
pairs of  
Gorenstein singularities and associated LG superconformal field theories
which have {\it a priori}\/ connections to the finite subgroups and
the modular invariants.  Unlike the situation for $SU(2)$, such an
attempt to find SCFT descriptions of Gorenstein singularities is 
hindered by the complexity of singularity structures and their
desingularizations.  Furthermore, we have seen that there is no direct
correspondence between LG superpotentials and isolated quotient
singularities; even though LG superpotentials have isolated
singularities, isolated quotient singularities 
cannot be represented as hypersurfaces and thus cannot be LG
superpotentials.  We have been thus led to more indirect methods of
analyzing the possible connections between LG theories and Gorenstein
orbifolds.   We have chosen the  $SU(3)_n/SU(2)\times
U(1)$ KS model as our particular LG theory and searched for evidences
for a correspondence with a Gorenstein orbifold $\C^3/\Gamma$ by
studying the intersection homology of its resolution.  Among  many
subgroups $\Gamma\subset SU(3)$ that we have considered, we have found
some surprising matches between the resolution of $\cznzn$ and the
$SU(3)_n/SU(2)\times U(1)$ KS LG theory at level $n$.

Using different techniques of desingularization
generally leads to topologically inequivalent Calabi-Yau manifolds with 
different Hodge numbers.  Furthermore, we know that the KS LG theories
describe deformations and not resolutions of Gorenstein orbifolds.  At
first sight, one might expect some kind of mirror symmetry between the
resolved $\cznzn$ and the LG orbifold.  
The complete resolution of the singularity produces a 
rigid Calabi-Yau with no complex deformations.  
On the other hand, one 
can embed the orbifold in $\C^4$ by the equation $xyz=w^{n+3}$ and try to 
deform the algebraic equation to remove the singularities, in which case 
the Hodge numbers are completely different from the resolution picture. 
In fact, with judicious choices of deformations, it is possible to 
produce a CY manifold with no second cohomology.
After all, certain rigid manifolds are known to have 
as their mirrors Landau-Ginzburg theories, whose physical modes have a 
natural interpretation as complex deformations of the superpotential but 
which have no explicit, geometric K\"{a}hler modes.  Thus, it may
appear to be not all 
inconceivable that there exist Landau-Ginzburg mirrors of the
complete or partial resolutions of the Gorenstein orbifolds.  A
careful investigation, however, shows that there are problems with
this picture.  In particular, the K\"{a}hler classes of the resolved
manifolds correspond to CFT moduli fields, but the chiral ring
in general consists of relevant, marginal, and irrelevant
operators\footnote{In LG theories, there exists a unique state whose
$U(1)$ charge is equal to $\hat{c}$.  Since $\hat{c}=2n/(n+3)$ for the
$SU(3)_n/SU(2)\times U(1)$ KS model, there will always be irrelevant
chiral ring elements for $n>3$.}.  They are thus not mirror pairs.

What is more likely to be true is that
the KS theory actually describes some
deformation of $\cznzn$ which is quite different from the picture that
traditional string theory techniques yield.  Thus, besides the two
known string theory Calabi-Yau 
desingularizations, with and without discrete
torsion, among thousands of other  ones allowed by
mathematics, perhaps certain KS theories can provide us with exactly
solvable descriptions of new desingularizations of Gorenstein
orbifolds.

We have dismissed many issues in this paper, mainly because not many
things are known about the interpretation of KS
LG theories as describing non-compact Calabi-Yau manifolds.
It would be interesting to see whether the
full quantum cohomology of the resolved Gorenstein orbifold 
can be related to the chiral
ring of some KS LG model, but this subject is clearly 
beyond the scope of this paper. 
The reader may have noticed that, in our presentation, we have largely
ignored
the non-compact factor $SL(2,\R)/U(1)$ and the $U(1)$ projection onto
integral charges.  It is because the main focus of the paper lies in
classifying the WZW modular invariants using geometry and because the
relevant information is contained in the $SU(3)_n/SU(2)\times U(1)$
sector of the tensored theory\footnote{Furthermore, imitating the ideas of
\cite{Ooguri-Vafa}, we want to send the coefficient $\mu$ in
\eqr{eq:W_T} to zero and argue that the tensored theory describes the
degenerating limit of an affine variety defined by the
superpotential.}.  
Despite our effort, there is yet no explanation of
why we should {\it a priori} expect the
finite subgroups to classify certain modular invariants of conformal
systems beyond the $SU(2)$ case.

Among the questions that we have ignored are: What would the
integrable deformations of the KS theory correspond to in terms of
geometry?  Can we include $D$-branes in our study?  What KS fields 
correspond to the non-compact ruled surfaces in the resolution; are
they related to the $SL(2,\R)/U(1)$ sector?

Finally, note that
 the Gorenstein orbifolds of the form $\C^3/\Z_n\times\Z_n$ have also appeared in the
context of AdS/CFT correspondence \cite{Morrison}.  The
``non-spherical near
horizon'' geometry of a $D3$-brane located at the origin of $\C^3/\Z_n\times\Z_n$,
for $n=2,3$,
has been shown to be given by $U(1)$ bundles over certain del Pezzo
surfaces.  It is also known that the $SU(2)/U(1)$ Kazama-Suzuki model
is closely related to the near horizon geometry of NS5-branes.  It
would be worth studying whether a similar picture exists for other
Grassman Kazama-Suzuki models.  Furthermore, regarding the modern
string compactifications on AdS backgrounds,
 we are reminded of the situation in
1980's where the validity of Calabi-Yau compactifications was
justified to some extent by replacing geometric compactifications with
algebraic counterparts using exactly solvable $N=2$ minimal models
\cite{Gepner}.   Similarly, it would be very interesting
to see whether there exist exactly solvable superconformal
 systems\footnote{We suspect that 
 RR-backgrounds may cause difficulties.}
describing the AdS compactifications.
We hope that our present study
may well be pertinent to such directions of pursuit.

%%%%%%%%%%%%%%%%%
\vspace{1cm} 

\noindent
{\bf Acknowledgments}

\noindent
We would like to thank C. Vafa for many valuable discussions and
suggestions.  We  also gratefully acknowledge Y.-H.E. He, D. Morrison, G. Tian, R. Vakil,
and Stephen S.-T. Yau
for discussions and Y.S. Song for comments on the preliminary
version of this paper.

%%%%%%%%%%%%%%%%%
%  Appendix
%%%%%%%%%%%%%%%%%
\newpage
\setcounter{equation}{0}
\appendix
\section{Appendix}
In this appendix, we determine which abelian quotient singularities
are isolated.  From Theorem~\ref{theorem:yau}, we can prove the
following simple results:

\begin{corollary}\label{cor:1}  The orbifold $\C^3/\Z_k, \Z_k\subset SU(3)$,
 has only an isolated singularity if and only
if $\Z_k = \langle\frac{1}{k}(\alpha_1, 
\alpha_2, \alpha_3) \rangle$ has GCD$(k,\alpha_i) = 1, \forall i$.
 In particular, $\C^3/ \Z_k$
always has non-isolated singularities for $k$ even.
\end{corollary}
	\noindent {\sc Proof:} 
 Let $\Z_k = \langle \omega \rangle$, where
$\omega = \frac{1}{k}(\alpha_1, \alpha_2, \alpha_3).$  It is easy to
 see that any non-trivial
 element $\omega^n\in\Z_k$ does not have an eigenvalue $1$ if and
 only if GCD$(k,\alpha_i)=1, \forall i$, and the first claim thus follows
 from Theorem~\ref{theorem:yau}.
Now, since $\Z_k \subset 
SU(3)$, the $\alpha_i$'s satisfy the condition $\alpha_1 +\alpha_2+
\alpha_3 =0\ (\mbox{mod } k)$, which implies that for $k$ even,
at least one $\alpha_i$, say $\alpha_1$, also has to be
even.  Let $m =
\mbox{GCD}(\alpha_1,k)\geq 2$.  Then, the action of the non-trivial element
$\omega^{k/m}$ on $\C^3$ fixes the first coordinate and thus produces
a non-isolated singularity along this axis.  \hfill\qed

\vspace{5mm}

Furthermore, a product of two cyclic groups satisfying the above
conditions yields 
an isolated singularity if and only if their orders are coprime:

\begin{corollary}\label{cor:kk'} Let  $\Z_k =\langle\frac{1}{k}(\alpha_1,
\alpha_2, \alpha_3)\rangle\subset SU(3)$ and $\Z_{k'} =\langle\frac{1}{k'}(\alpha'_1,
\alpha'_2, \alpha'_3)\rangle\subset SU(3)$, where $0<\alpha_i<k$ and $0<\alpha'_i <k'$.
Assume
that $k, k'$ are odd and that  GCD$(k,\alpha_i) = \mbox{GCD}(k',
\alpha'_i)=1,\, \forall i$.  Then,
$\C^3/\Z_k
\times \Z_{k'}$ has only isolated singularities if and only if $k$ and $k'$ are
coprime.
\end{corollary}
\noindent {\sc Proof:} Without a loss of generality, assume that
$k'<k$.  In the diagonal basis, we can represent
any non-trivial elements 
$g\in\Z_k$ and $g'\in\Z_{k'}$ as $g=\frac{1}{k} (a_1,a_2,a_3)$ and
$g'=\frac{1}{k'} (b_1,b_2,b_3)$, for some integers $0<a_i<k$ and $0<b_i<k'$.  We
see that $gg'$ has an eigenvalue $1$ if and only if
	\beq 
	\frac{a_i}{k} + \frac{b_i}{k'} =1 	\label{eq:coprime}
	\eeq
or equivalently,
	\beq
	a_i = \frac{k (k' -b_i)}{k'} 	\label{eq:a_i}
	\eeq
for some $i$.  Now, assume that GCD$(k,k')=1$.  Then, \eqr{eq:a_i}
tells us that in order for
$a_i$ to be an integer, $k'$ has to divide $k'-b_i$, which is
impossible.  

Conversely, suppose that GCD$(k,k')=c > 1$, such that $k=cm$ and
$k'=cn$ for some positive coprime integers $m,n$.  But, because we
have assumed that 
GCD$(k,\alpha_i) = \mbox{GCD}(k',\alpha'_i)=1\ \forall i$,
 there will be elements $g= \frac{1}{k} (a_1,a_2,a_3)$ and
$g'=\frac{1}{k'} (b_1,b_2,b_3)$
 for any
$d, 1\leq d <c$,
such that $a_1 = dm$ and $b_1= n(c-d)$.  Then, we have
	\[ 
	\frac{a_1}{k} + \frac{b_1}{k'}= \frac{a_1n + b_1m}{cmn} = 1
	\]
and thus, $gg'$ has an eigenvalue $1$.   \hfill\qed

\vspace{5mm}
Corollary~\ref{cor:kk'} just means that $\Z_k \times \Z_{k'} \cong
\Z_{kk'}$, where  $\Z_{kk'}$ must 
satisfy the conditions of Corollary~\ref{cor:1}.
Corollary~\ref{cor:3} now follows.

%%%%%%%%%%%%%%%%%%%%%#############################################
%  	Reference
%%%%%%%%%%%%%%%%%%%%%#############################################
\newpage
\nc{\bi}{\bibitem}

\end{document}